# Flattening-off of droplet bouncing trend under high ambient gas pressures

Chenwei Zhang(张宸玮)[a,b], Zhenyu Zhang(章振宇)[a,b*], Peng Zhang(张鹏)[c*], Jinhui Zhou(周晋辉)[a,b] and Changlu Zhao(赵长禄)[a,b]

[a] *School of Mechanical Engineering, Beijing Institute of Technology, Beijing 100081, P. R. China*

[b] *National Key Laboratory of Science and Technology on Multi- perch Vehicle Driving Systems, Beijing Institute of Technology, Beijing 100081, P. R. China*

[c] *Department of Mechanical Engineering, City University of Hong Kong, Hong Kong 999077, P. R. China*

**Abstract**: It was previously observed that colliding liquid droplets in a gaseous medium tend to bounce off at elevated gas pressure up to about 12 atm. In this letter, we extended the droplet collision experiment to up to 41 atm for the first time and reported a noticeable discovery that the tendency is flattened off at higher pressures. The colliding droplets stop bouncing but start to coalesce beyond a critical Weber number, which increases with pressure but tends to a limit value at 21 atm and above. A scaling analysis taking into account the gas-film dynamics, the rarefied gas effects, and van der Waals force well correlates with the experimental discovery.

\* Corresponding author
*E-mail address:* zhenyu.zhang@bit.edu.cn (Zhenyu Zhang)
penzhang@cityu.edu.hk (Peng Zhang)

Droplet-droplet collision is a common phenomenon widely existing in nature and industrial applications [1-4]. It gained long-time attention in meteorological research due to its important relevance in the formation of raindrops, and the outcomes of coalescence and separation have been discovered [5-7]. With its relevance in spray combustion being recognized in liquid-fueled combustion engines, the collision between alkane droplets has been experimentally studied and droplet bouncing has been observed, particularly at elevated pressures [8-11]. Among these experiments, different collision outcomes were often mapped in a nomogram whose coordinate axes are the Weber number ($We = 2\rho_l U^2 R_s/\sigma$) and the collision parameter ($B = b/(R_s + R_L)$). Here, $\rho_l$, $R_s$, $R_L$, $\sigma$, $U$ are liquid density, radii of small and large droplets, surface tension, and relative velocity, respectively, and $b$ is the projection of two liquid droplets' centerline perpendicular to the relative velocity. Delineating the boundaries between the collision outcomes regime in the nomogram has been a focus of research because of its significance to spray modeling [4,12-14]. Previous studies have demonstrated that the boundaries are subject to liquid composition, viscosity, droplet size ratio, and many other relevant parameters [15-22]. Additionally, the variation in the ambient environments such as temperature and pressure also substantially influences the collision outcomes [11,23-26].

The promoting effect of increasing ambient pressure on droplet bouncing has been confirmed, which is thought to be associated with the weakening of the rarefied gas effect due to the elevation in ambient pressure [11,27]. Although the finding offers valuable insights for enhancing the existing droplet collision models in Lagrangian simulations [28-31], the underlying physics has not been well understood, especially how the rarefied gas effect acts in the gas film under different ambient pressure remains to be explored since the maximum pressure achieved in previous experiments was only 12 atm.

To investigate the influence of ambient pressure on the outcomes of droplet collisions, particularly the behavior under high pressures that is of relevance to real engine conditions. A novel experimental system was established recently and it consists of four main components: the high-pressure chamber, droplet generation system, gas drive system, and image acquisition system as shown in Fig. 1(a). Two droplets will generated by two symmetrically arranged microvalves (Gyger SMLD 300G) which are fixed on the four-dimensional positioners. As shown in Fig. 1(c), the operational performance of the microvalves can be controlled by varying the frequency and time of the electrical signals, and the size of the generated droplets can be controlled within the range of 200-300 micrometers at the appropriate pulse width. It must be emphasized that inducing such tiny droplets to collide within the high-pressure chamber is a quite challenge, which is overcome by the precise positioning and orientation of the microvalves. To ensure the high-pressure chamber sealing while achieving accurate control of the four-dimensional positioners, aerial insertion was used for the signal transmission. As shown in Fig. 1(b), the movement of the four-dimensional positioners in the Y-direction enables droplet collision within a singular, unified plane, specifically, the focal plane of the camera as far as possible. Movements in other directions are employed to meticulously manage the relative positioning of droplets on either side, guaranteeing the occurrence of droplet collisions by exercising precise control with an accuracy of 2 microns. All the movements of different directions can be finely controlled by adjusting the signal pulse width and altering the forward or backward direction. The entire collision process can be recorded by a high-speed camera (Photron NOVA Fastcam S12) with 25000 frames/s. To capture high-resolution images of the droplets, which have a small size ranging from 200 to 300 microns, a microlens (MAZ12.0X-LZ) was utilized at its maximum magnification of 7.2X with the spatial resolution of 5.14 $\mu m/pixel$. In contrast to droplet collision in the atmosphere, the installation of a camera directly above the colliding droplets is considerably challenging. It will inevitably result in a decrease in the accuracy of estimating collision parameters although the four-dimensional positioners are capable of modifying the fore and aft positions of two liquid droplets [32], since it is impossible to eliminate droplet misalignment in the shooting direction through experimental methods. Hence, an image processing technique employing droplet edge detection is used to mitigate the computational error arising from this scenario to the greatest extent possible. More details about the image processing method are provided in the Supplemental Material.

The experimental system outlined above enabled the recording of the droplet collision process at ambient pressures reaching up to 41 atm. As shown in Fig. 2, four different collision outcomes including bouncing, coalescence, reflective separation, and stretching separation were recorded. It is known that droplet collision with a sufficiently small Weber number tends coalesce [11,33], but the present droplet generation system could not produce such coalescence. Furthermore, the outcomes of droplet collision can be roughly divided into two categories, depending on whether interface fusion occurs after temporary collision, which are bouncing and others, due to the fact that separation is a phenomenon occuring after droplet coalescence. And therefore, a prerequisite for predicting the liquid droplet collision outcomes is the accurate determination of the boundaries between bouncing and coalescence.

Fig. 3 presents data on droplet collisions across various ambient pressures up to 41 atm. As the ambient pressure increases, the bouncing regime exhibits a persistent expansion. This observation aligns with prior experimental findings [11,26] and theoretical analysis [11,27], which show the elevation in ambient pressure facilitates droplet bouncing. This is because a higher ambient pressure will hamper the gas film drainage, so the two droplets cannot be adequately close to each other to trigger the van der Waals force. Ultimately, droplet coalescence does not occur and the droplets bounce off by the surface tension force.

The most significant promotion of droplet bouncing by increasing pressure occurs when the pressure

($p$) increases from 3 atm to 11 atm, resulting in a substantial increase in the critical Weber number ($We_{cr}$) between droplet bouncing and coalescence, as shown in Fig. 3(a) to Fig. 3(e). It is striking that $We_{cr}$ does not exhibit a significant increase with further increasing the ambient pressure from 11 atm to 41 atm, as shown in Fig. 3(e) to Fig. 3(h). This was never discovered in any previous studies because they were limited to the ambient pressure of up to 12 atm.

To quantitatively understand this phenomenon, we must precisely demarcate various regimes of droplet collisions. Instead of using empirical polynomial fitting, we adopted a physical model, $We_{cr} = 8(\phi' - 3)/\chi(1 - B^2)$, originally proposed by Estrade et al.[8] and improved by many studies [4,12,28,29,34,35]. The parameter $\phi' = 2\left(\frac{6}{\phi} - 2\right)^{-\frac{1}{3}} + \left(\frac{6}{\phi} - 2\right)^{\frac{2}{3}}$ is derived by Karrar et al.'s [12] from the mass conservation. $\chi$ is a quantity related to the effective contact volume of two droplets, $\chi = 1 - 0.25(2 - \tau)^2(1 + \tau)$ for $\tau > 1$, and $\chi = 0.25\tau^2(3 - \tau)$ otherwise, where $\tau$ is defined by $(1 - B)(1 + \Delta)$ and is reduced to $2(1 - B)$ when two droplets are identical. The shape factor $\phi$ quantifying the droplet deformation strongly depends on the impact parameter $B$ [12], so it is fitted as a function $aB^b + c$, where $a, b, c$ are fitting parameters and $B$ is the collision parameter. Then we conducted parameter optimization for these variables by employing the accuracy which is the proportion of accurately classified data points to the total number of data points as the index. To ensure the precision of the boundary, the potential errors in the experimental data that may arise from the image processing procedure have also been taken into account. Therefore, the boundary that used to separate bouncing and coalescence has transformed into a collaborative group. As shown in Fig. 3, any boundary line within the blue area, situated between the upper and lower boundaries, can be represented as the dividing line that exhibits the highest level of accuracy. More details about the boundary determination and error analysis are provided in the Supplemental Material.

To evaluate the critical Weber numbers under various collision parameters simultaneously, $(1 -$

$B^2)\sqrt[3]{\chi}$ was multiplied to $We$ and the equivalent Weber number $We^*$ was obtained [8]. Then with boundary delineation methodology, we successfully identified the variation pattern of $We_{cr}^*$ in relation to ambient pressures, as shown in Fig. 4. It is evident that increasing the ambient pressure promotes droplet bouncing. However, $We_{cr}^*$ ceases to exhibit significant increments beyond 21 atm. To shed light on this observation, we extended the minimum thickness theory proposed by Huang et al. [36] by including the rarefied gas effects. According to Huang et al. [36],

$$\frac{24 \cdot Oh_{g.l}\sqrt{We_{cr}^*}}{We_{cr}^* + 4 - 8Oh_l\sqrt{We_{cr}^*}} = 21.1 \sqrt[3]{\frac{A^*}{24\pi}} \tag{1}$$

where $Oh_l = \mu_l/\sqrt{\rho_l D\sigma}$, $Oh_{g.l} = \mu_g/\sqrt{\rho_l D\sigma}$, $A^* = A_H/\sigma D^2$, $\mu_l$ is gas viscosity (water in the experiment), $\mu_g$ is gas viscosity (nitrogen in the experiment), $D$ is droplet diameter, and $A_H$ is the Hamaker constant.

We will subsequently modify the equation to incorporate the influence of ambient pressure by considering the rarefied gas effects. According to Zhang and Law's theory [27], the flow within the lubrication layer taking into account the rarefied flows can be expressed:

$$p_g = \frac{3\mu_g/\Delta(Kn)}{h^3}(r^2 - a^2)\left(\frac{dh}{dt} + 2\kappa h\right) \tag{2}$$

where the additional correction factor $\Delta(Kn)$ is given by,

$$\Delta(Kn) = 1 + 6.0966Kn + 0.965Kn^2 + 0.6967Kn^3 (Kn < 1) \tag{3a}$$

$$\Delta(Kn) = 8.7583Kn^{1.1551}(Kn \geq 1) \tag{3b}$$

$Kn = \lambda/h$ is the Knudsen number, which varies with the gas film thickness $h$. Through the implementation of this method, Li [37] seamlessly incorporated the rarefied gas effect into the N-S equation, enabling precise predictions of droplet bouncing and coalescence after collision in their simulations. Chubynsky et al. [16] also achieved the description of the rarefied gas effect by dividing viscosity by $\Delta(Kn)$ in their simulation as:

$$\Delta(Kn) = 1 + 6.88Kn + \frac{6Kn}{\pi}\ln(1 + 2.76Kn + 0.127Kn^2) \qquad (4)$$

And more details about the comparison of Equ. (3) and Equ. (4) are shown in Supplemental Material. We first solve Eq. (1) to obtain $We_{cr}^*$ as:

$$We_{cr}^{*\frac{1}{2}} \sim \frac{\mu_g}{\sqrt{\rho_l D \sigma}} \sqrt[3]{\frac{24\pi}{A^*}} \qquad (5)$$

The modified critical Weber number with rarefied gas effects can be represented as:

$$We_{cr}^{*\frac{1}{2}} \sim \frac{\mu_g/\Delta(Kn)}{\sqrt{\rho_l D \sigma}} \sqrt[3]{\frac{24\pi}{A^*}} \qquad (6)$$

Therefore, the variation of $We_{cr}^*$ under different ambient pressures is determined by the Knudsen number as the Hamaker constant, viscosity, surface tension of liquids, and gas viscosity exhibit minimal variation in response to changes in pressure. By using Equ. (3a) up to $O(Kn)$, we have

$$We_{cr}^{*-\frac{1}{2}} \sim \frac{\sqrt{\rho_l D \sigma}}{\mu_g} \sqrt[3]{\frac{A^*}{24\pi}} (1 + 6.0966Kn) = (a_0 + a_1 \tilde{p}^{-1}) \qquad (7)$$

where $\tilde{p}$ is dimensionless pressure which equals to the ratio of ambient pressure to atmospheric pressure $p/p_0$. Apparently, $a_0$ can be estimated by $a_0 = \lim_{p \to \infty} We_{cr}^{*-\frac{1}{2}} \approx 0.23$ and $a_1$ can be estimated by $a_1 = We_{cr}^{*-\frac{1}{2}}(p_\infty) - a_0 = 0.3$, where we have used $h = 0.001D$ as the characteristic gas film thickness [36]. The comparison of the prediction of Equ. (7) with the experimental data are shown in Fig. 4.

In summary, we investigated the influence of ambient pressure on droplet collisions, with a particular focus on the delineation between droplet bouncing and other outcomes. The experimental findings align with prior research, indicating that increased pressure facilitates droplet bouncing after collision. However, slight deviations appear when the ambient pressure reaches a sufficiently high level. To elucidate this disparity, the boundary between droplet bouncing and other outcomes has been determined first based on the Estrade et al.' model [8] and Karrar et al.'s modification [12]. And the discovery of the phenomenon wherein the influence of ambient pressure on droplet collision outcomes gradually degenerates has been made first. Then the theoretical analysis based on Huang et al. [36] and Zhang and Law [27] was conducted. Finally, the prediction of critical Weber numbers between droplet bouncing and coalescence with variations in ambient pressure was completed.

This work was supported by the National Natural Science Foundation of China (Grant No. 51806013 and No. 52176134). The work at the City University of Hong Kong was additionally supported by grants from the Research Grants Council of the Hong Kong Special Administrative Region, China (Project No. CityU 15222421 and CityU 15218820). The authors are grateful to Dr. Tao Yang and Dr. Yicheng Chi for their insightful advice for the data processing and to Dr. Ning Wang for his suggestion for the theoretical analysis.

# Reference:

[1] W. Bauer, G. F. Bertsch, and H. Schulz, Physical Review Letters **69**, 1888 (1992).
[2] H. Wu, F. Zhang, and Z. Zhang, Physics of Fluids **33**, 013317 (2021).
[3] C. Zhang, Z. Zhang, K. Wu, X. Xia, and X. Fan, Physics of Fluids **33**, 093311 (2021).
[4] Z. Zhang, Y. Chi, L. Shang, P. Zhang, and Z. Zhao, International Journal of Heat and Mass Transfer **102**, 657 (2016).
[5] N. Ashgriz, Journal of Fluid Mechanics **221**, 183 (1990).
[6] P. R. Brazier-Smith, S. G. Jennings, and J. Latham, Proceedings of the Royal Society A: Mathematical, Physical and Engineering Sciences **326**, 393 (1972).
[7] R. Gunn, Science **150**, 695 (1965).
[8] J. P. Estrade, H. Carentz, G. Lavergne, and Y. Biscos, International Journal of Heat and Fluid Flow **20**, 486 (1999).
[9] Y. J. Jiang, A. Umemura, and C. K. Law, Journal of Fluid Mechanics **234**, 171 (1992).
[10] K. L. Pan, C. K. Law, and B. Zhou, Journal of Applied Physics **103**, 064901 (2008).
[11] J. Qian and C. K. Law, Journal of Fluid Mechanics **331**, 59 (1997).
[12] K. H. Al-Dirawi and A. E. Bayly, Physics of Fluids **31**, 027105 (2019).
[13] S. L. Post and J. Abraham, International Journal of Multiphase Flow **28**, 997 (2002).
[14] M. Sommerfeld, International Journal of Multiphase Flow **27**, 1829 (2001).
[15] G. Brenn and A. Frohn, Experiments in Fluids **7**, 441 (1989).
[16] M. V. Chubynsky, K. I. Belousov, D. A. Lockerby, and J. E. Sprittles, Physical Review Letters **124**, 084501 (2020).
[17] G. Finotello, R. F. Kooiman, J. T. Padding, K. A. Buist, A. Jongsma, F. Innings, and J. A. M. Kuipers, Experiments in Fluids **59**, 1 (2018).
[18] C. Gotaas, P. Havelka, H. A. Jakobsen, H. F. Svendsen, H. Matthias, N. Roth, and B. Weigand, Physics of Fluids **19**, 102106 (2007).
[19] X. Jia, J. C. Yang, J. Zhang, and M. J. Ni, International Journal of Multiphase Flow **116**, 80 (2019).
[20] N. Nikolopoulos, G. Strotos, K. S. Nikas, and G. Bergeles, International Journal of Heat and Mass Transfer **55**, 2137 (2012).

[21] C. Rabe, J. Malet, and F. F. euillebois, Physics of Fluids **22**, 047101 (2010).
[22] C. Tang, P. Zhang, and C. K. Law, Physics of Fluids **24**, 695 (2012).
[23] N. Ashgriz and P. Givi, International Journal of Heat and Fluid Flow **8**, 205 (1987).
[24] K. D. Willis and M. E. Orme, Experiments in Fluids **29**, 347 (2000).
[25] K. Willis and M. Orme, Experiments in Fluids **34**, 28 (2003).
[26] L. M. Reitter, M. Liu, J. Breitenbach, K. L. Huang, and C. Tropea, in *Ilass-european Conference on Liquid Atomization & Spray Systems*2017).
[27] P. Zhang and C. K. Law, Physics of Fluids **23**, 042102 (2011).
[28] C. Hu, S. Xia, C. Li, and G. Wu, International Journal of Heat and Mass Transfer **113**, 569 (2017).
[29] K. G. Krishnan and E. Loth, International Journal of Multiphase Flow **77**, 171 (2015).
[30] Z. Zhang and P. Zhang, Physics of Fluids **29**, 103306 (2017).
[31] X. Wei and Z. Zhang, Computers & Fluids **246**, 105621 (2022).
[32] K. L. Huang, K. L. Pan, and C. Josserand, Physical Review Letters **123**, 234502 (2019).
[33] A. Gopinath and D. L. Koch, Journal of Fluid Mechanics **454**, 145 (2002).
[34] M. Sommerfeld and M. Kuschel, Experiments in Fluids **57**, 1 (2016).
[35] M. Sommerfeld and L. Pasternak, International Journal of Multiphase Flow **117**, 182 (2019).
[36] K. L. Huang and K. L. Pan, Journal of Fluid Mechanics **928**, 1, A7 (2021).
[37] J. Li, Physical Review Letters **117**, 214502 (2016).

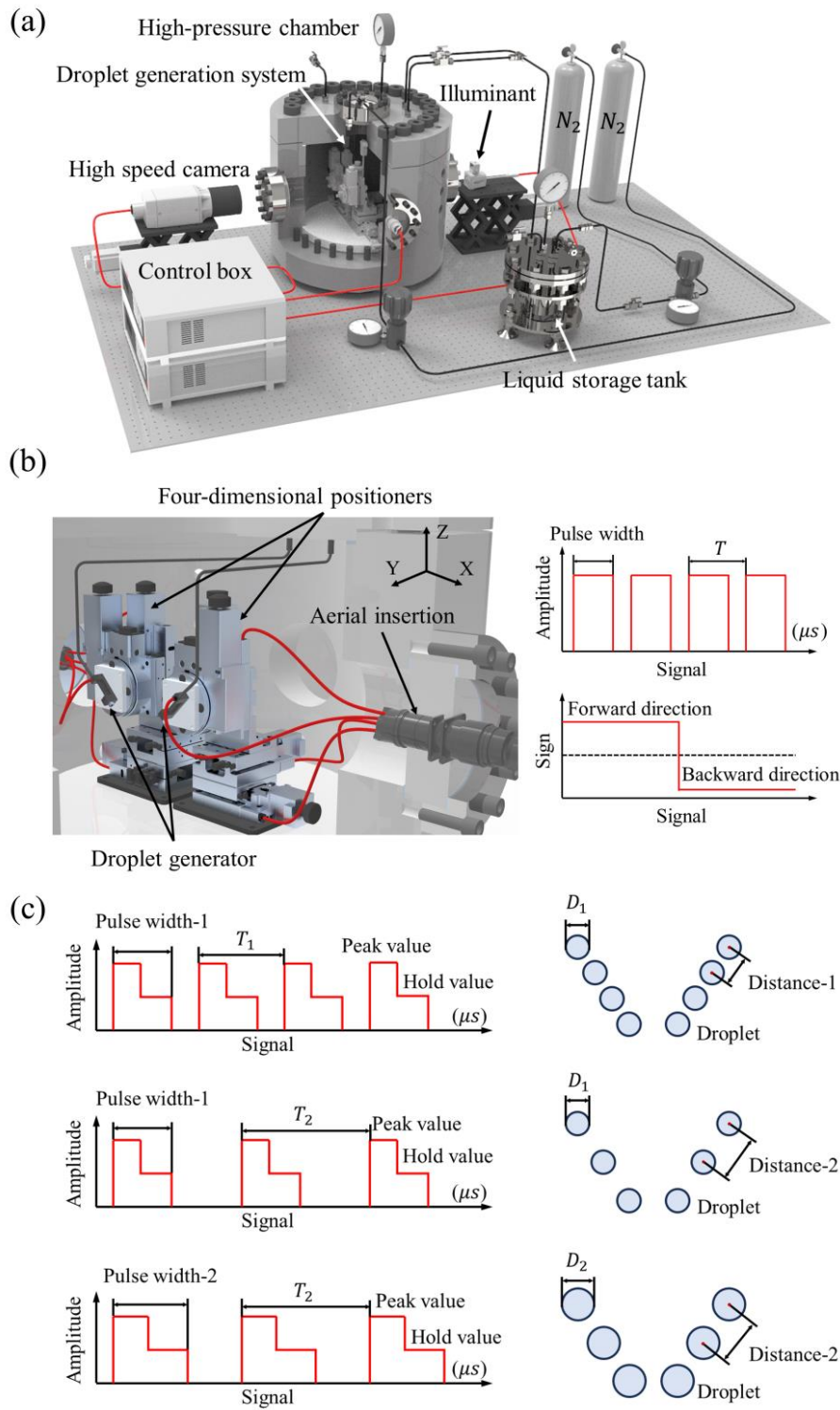

Figure 1. (a) A novel experimental apparatus for droplet collision at elevated pressures up to 41 atm, (b) the droplet generation system within the high-pressure chamber and the control signal, and (c) the microvalve control signals and varying control signals on droplet formation.

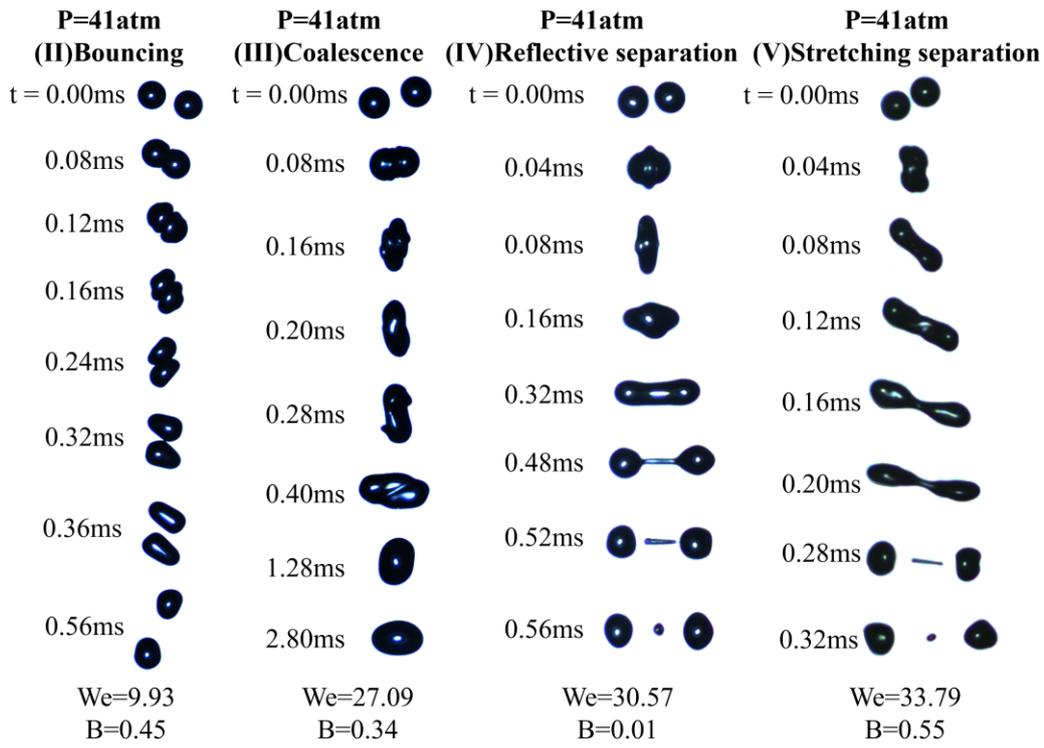

Figure 2. Different outcomes of droplet collision at 41 atm.

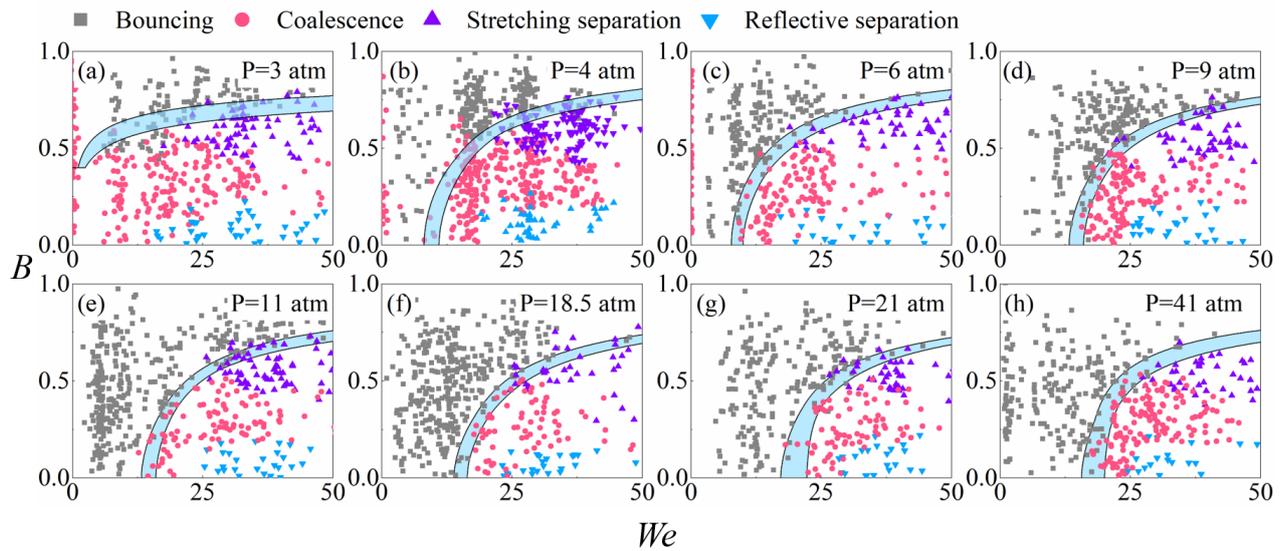

Figure 3. *We-B* nomograms of droplet collision outcomes under various ambient pressures. The blue stripes are the boundaries (with error bars) separating the bouncing regime from other regimes.

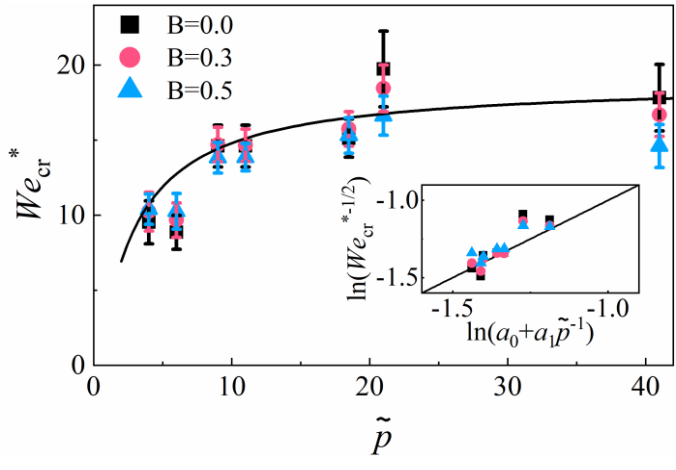

Figure 4. The variation of the defined critical Weber number in response to changes in ambient pressure.

Supplementary Material

of

Flattening-off of droplet bouncing trend under high ambient gas pressures

Chenwei Zhang(张宸玮)[a,b], Zhenyu Zhang(章振宇)[a,b*], Peng Zhang(张鹏)[c*], Jinhui Zhou(周晋辉)[a,b] and Changlu Zhao(赵长禄)[a,b]

[a] *School of Mechanical Engineering, Beijing Institute of Technology, Beijing 100081, P. R. China*

[b] *National Key Laboratory of Science and Technology on Multi- perch Vehicle Driving Systems, Beijing Institute of Technology, Beijing 100081, P. R. China*

[c] *Department of Mechanical Engineering, City University of Hong Kong, Hong Kong 999077, P. R. China*

\* Corresponding author
*E-mail address:* zhenyu.zhang@bit.edu.cn (Zhenyu Zhang)
penzhang@cityu.edu.hk (Peng Zhang)

# I. MORE ON THE IMAGE PROCESSING METHOD

In the preliminary phase of the experiment, the camera screen will display two droplet streams that may not be clearly visible due to the lack of proper focusing. Then, the distance between the droplet streams and the camera will be adjusted by manipulating the Y-axis movement of the positioners until the clear edges of the droplets emerge. The fulfillment of droplet collision with various collision parameters can be achieved by adjusting the R-axis and the high-resolution droplet collision images can be obtained and are shown in Fig. S1 to Fig. S6.

The purpose of our research is to investigate the influence of ambient pressure on the outcomes of droplet collisions. Therefore, it is necessary to carefully consider the accurate estimation of droplet Weber numbers and collision parameters. As shown in Fig. S7(a), the images were initially subjected to binarization, with a threshold value of 0.5 being selected [1,2]. The determination of droplet collision is based on the counting of connected domains. In cases where two droplets do not collide, there will always be two connected domains in the image. However, when the droplets come into contact, the number of connected domains will decrease to one. We designate the moment at which the number of connected domains is reduced to one as the reference time ($t_0$). Images captured prior to this reference time will be utilized for the computation of Weber numbers and collision parameters. This is because any deformation occurring after droplet collision can introduce errors in the estimation of droplet size. When measuring droplet diameters, it was found that it is not feasible to identify two droplets with precisely identical sizes. Referring to the suggestion proposed by Ashgriz et al., droplets exhibiting size errors below 10% are deemed equivalent, and the average value of two droplet sizes is used for the subsequent calculation [3].

Droplet misalignment can be a source of collision parameter calculation errors as shown in Fig. S7(b). To mitigate this error, an image processing technique employing droplet edge detection is proposed. We first proceed to analyze the droplet collision images that have been identified as not misaligned under atmospheric conditions, and extract the values of the G channel in the RGB images as the G-channel image exhibits the sharpest edges compared to the other channel images as shown in Fig. S8 to Fig. S10. Then the gradient values in both the transverse and vertical directions of the droplets were determined, and the integral values in each direction were subsequently computed. The definitions of transverse and vertical directions are shown in Figure 7(c). As shown in Fig. S11, the integration deviation resulting from misalignment-induced gradient variations does not surpass 5%. And therefore, this value will serve as a criterion for assessing droplet misalignment.

In addition to this, we conducted a comparison of the time scale employed for the calculation of the Weber number and collision parameter during droplet collision. As shown in Fig. S12(a) and (b), when using a single time interval as the time scale to do the calculation, the obtained values will have a certain degree of fluctuation. This is due to the fact that shorter time interval will increase the computational uncertainty. And this scenario can also be observed under varying ambient pressures, as shown in Fig. S12(e) and (f). On account of this phenomenon, the Weber number and collision parameter are calculated from $t_{-1}$ to a relatively long time period before and the obtained values are plotted in Fig. S12(c), (d), (g), and (h). It is evident that the amplitude of data fluctuations decreases significantly as the time interval increases, whether discussing the Weber number or collision parameters. Based on this characteristic, a selection of five images was made prior to the collision. The collision data from $t_{-1} - t_{-2}$ to $t_{-1} - t_{-5}$ was then calculated and averaged to determine the droplet Weber and collision parameter, which means:

$$We = \frac{1}{4} \sum_{n=1}^{n=4} We_{n\Delta t} \quad (1)$$

$$B = \frac{1}{4} \sum_{n=1}^{n=4} B_{n\Delta t} \quad (2)$$

Finally, an error analysis was conducted on the data preprocessing method mentioned above. As shown in Fig. S13, it can be observed that although the calculation error may exhibit a slight increase with the rise in ambient pressure, the Weber number exhibits an error margin of no more than 1, and the collision parameter demonstrates an error margin of less than 0.015.

## II. MORE ON BOUNDARYIES BETWEEN DROPLET BOUNCING AND OTHER OUTCOMES

Due to the inherent inaccuracies in the droplet collision data obtained through current data processing method, there is inevitably a possibility of errors in determining boundaries. In order to ascertain the boundary range, we performed boundary determinations utilizing the maximum data error limit. As shown in Fig. S14, taking the data under the ambient pressure of 20 atm as an example. We conducted parameter optimization considering the original experimental data, as well as the experimental data moved along the direction 1 and 2 according to the error, which means $(We + \frac{1}{2}We_{error}, B - \frac{1}{2}B_{error})$ and $(We - \frac{1}{2}We_{error}, B + \frac{1}{2}B_{error})$. By modifying data within specified error limits, the shape factor, as determined by this method, comprehensively account for the errors inherent in the image processing procedure. And this can also provide a more accurate representation of the distinction between the droplet bouncing and other outcomes.

## III. MORE ON THE COMPARISON OF EQU. (3) AND EQU. (4)

The discrepancy between Zhang and Law's theory [4] and Chubynsky et al.'s theory [5] is very small due to their modification of viscosity both involving the power of the Knudsen number as shown in Equ. (3a) and Equ. (4) in the letter. And therefore, their theories can be further simplified up to $O(Kn)$,

$$\Delta(Kn) \approx 1 + 6.0966 Kn \qquad (S1)$$

$$\Delta(Kn) \approx 1 + 6.88 Kn \qquad (S2)$$

for the tiny changes between simplified forms and the original ones as shown in Fig. S15.

# Reference:

[1]   P. Zhang and B. Wang, Physics of Fluids **29**, 042102 (2017).

[2]   C. Zhang, Z. Zhang, K. Wu, X. Xia, and X. Fan, Physics of Fluids **33**, 093311 (2021).

[3]   N. Ashgriz, Journal of Fluid Mechanics **221**, 183 (1990).

[4]   P. Zhang and C. K. Law, Physics of Fluids **23**, 042102 (2011).

[5]   M. V. Chubynsky, K. I. Belousov, D. A. Lockerby, and J. E. Sprittles, Physical Review Letters **124**, 084501 (2020).

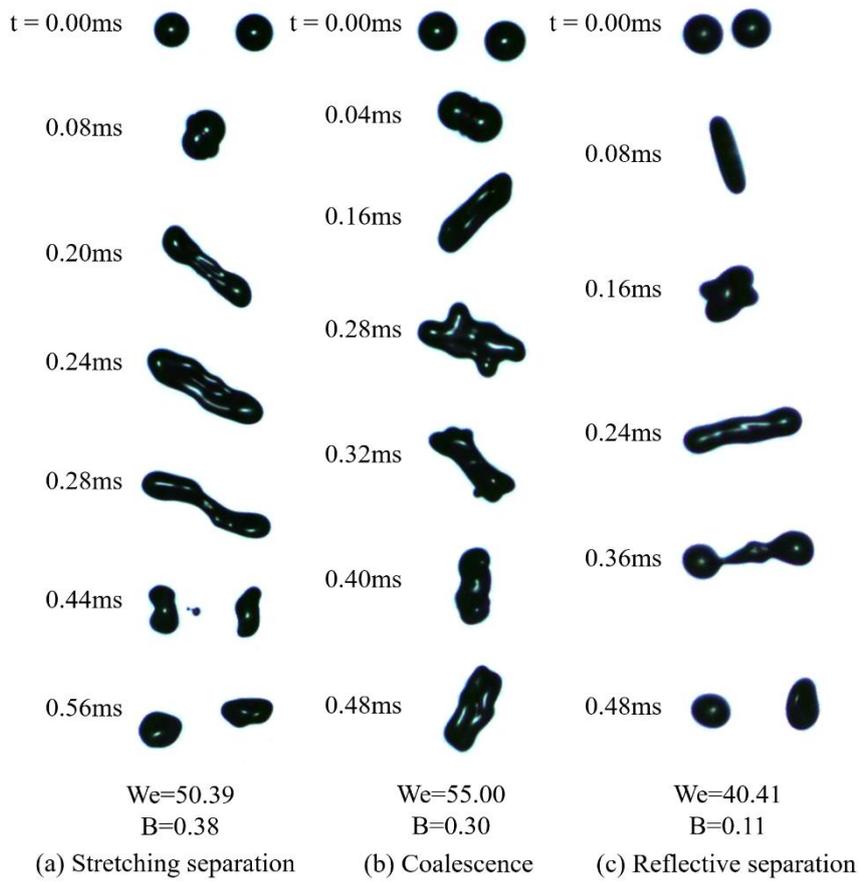

Figure S1. Different droplet collision outcomes under atmospheric pressure, (a) Stretching separation, (b) Coalescence, and (c) Reflective separation.

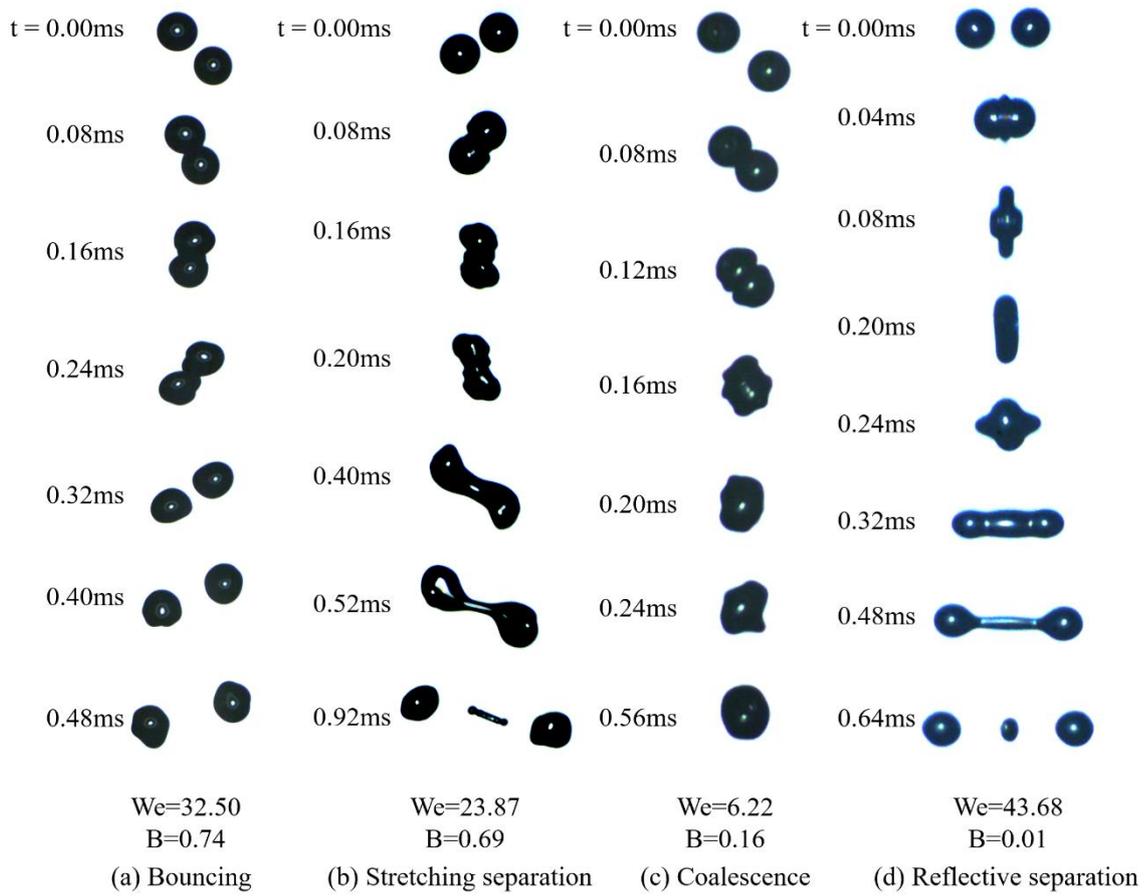

Figure S2. Different droplet collision outcomes under 3 atm, (a) Bouncing, (b) Stretching separation, (c) Coalescence, and (d) Reflective separation.

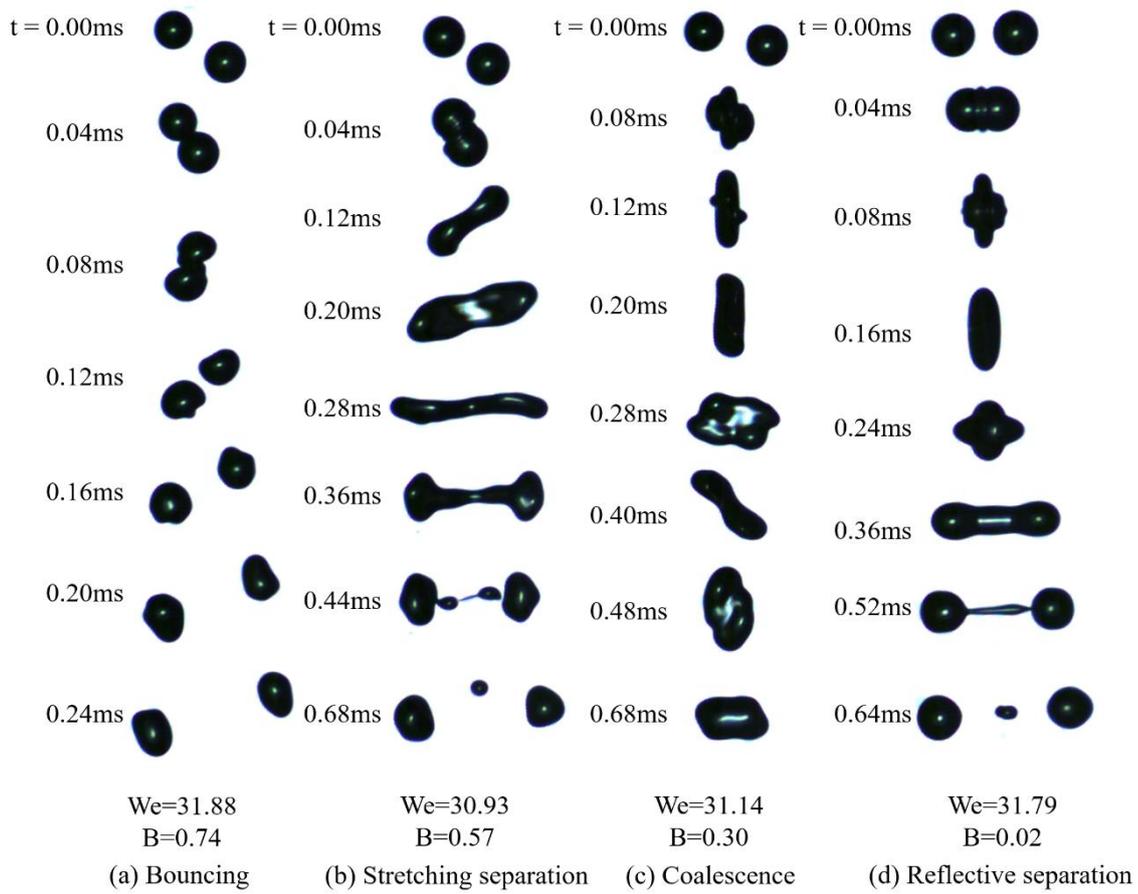

Figure S3. Different droplet collision outcomes under 4 atm, (a) Bouncing, (b) Stretching separation, (c) Coalescence, and (d) Reflective separation.

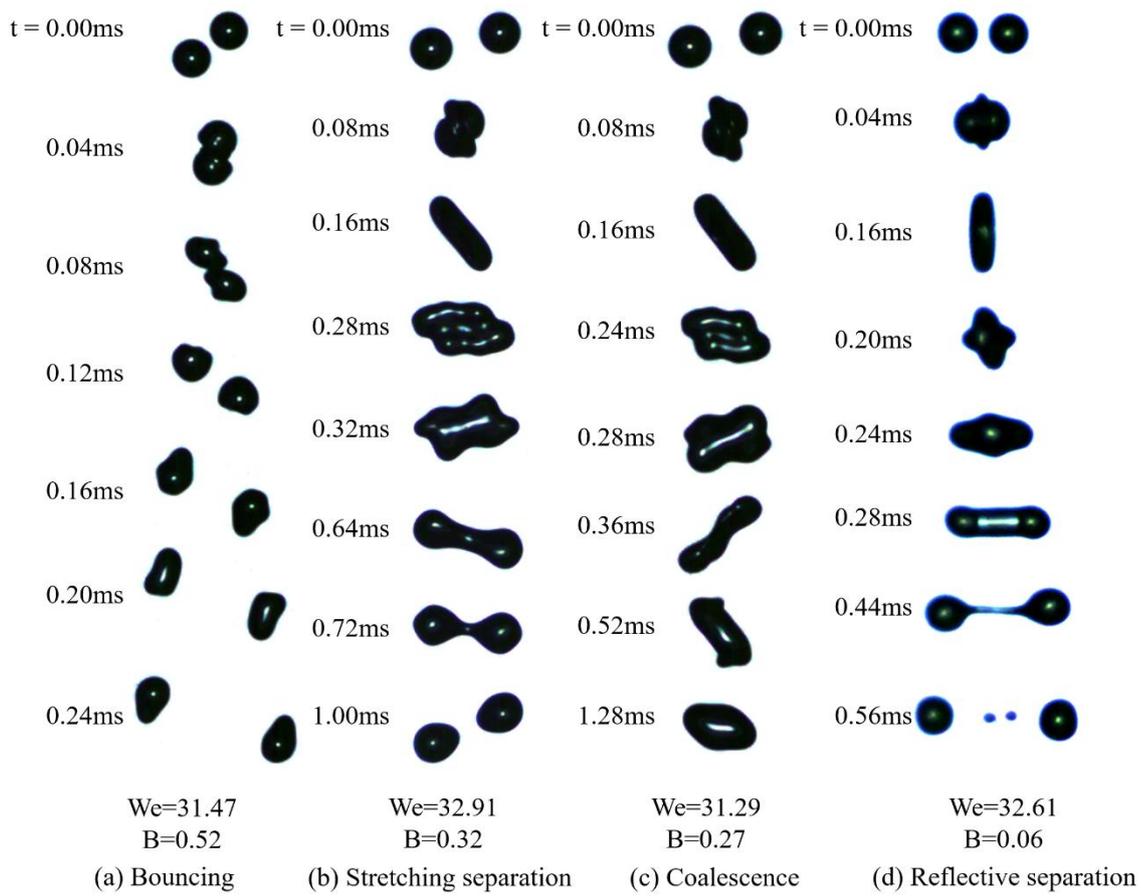

Figure S4. Different droplet collision outcomes under 9 atm, (a) Bouncing, (b) Stretching separation, (c) Coalescence, and (d) Reflective separation.

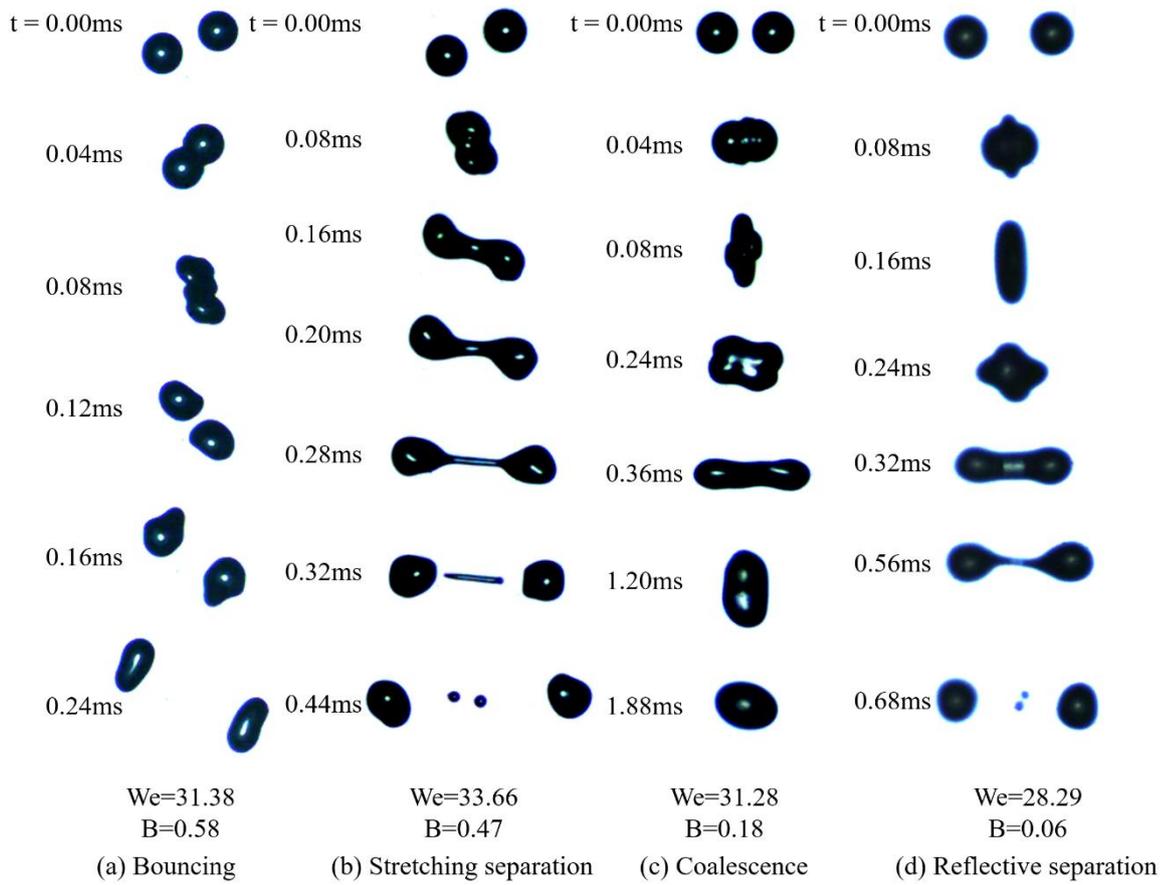

Figure S5. Different droplet collision outcomes under 11 atm, (a) Bouncing, (b) Stretching separation, (c) Coalescence, and (d) Reflective separation.

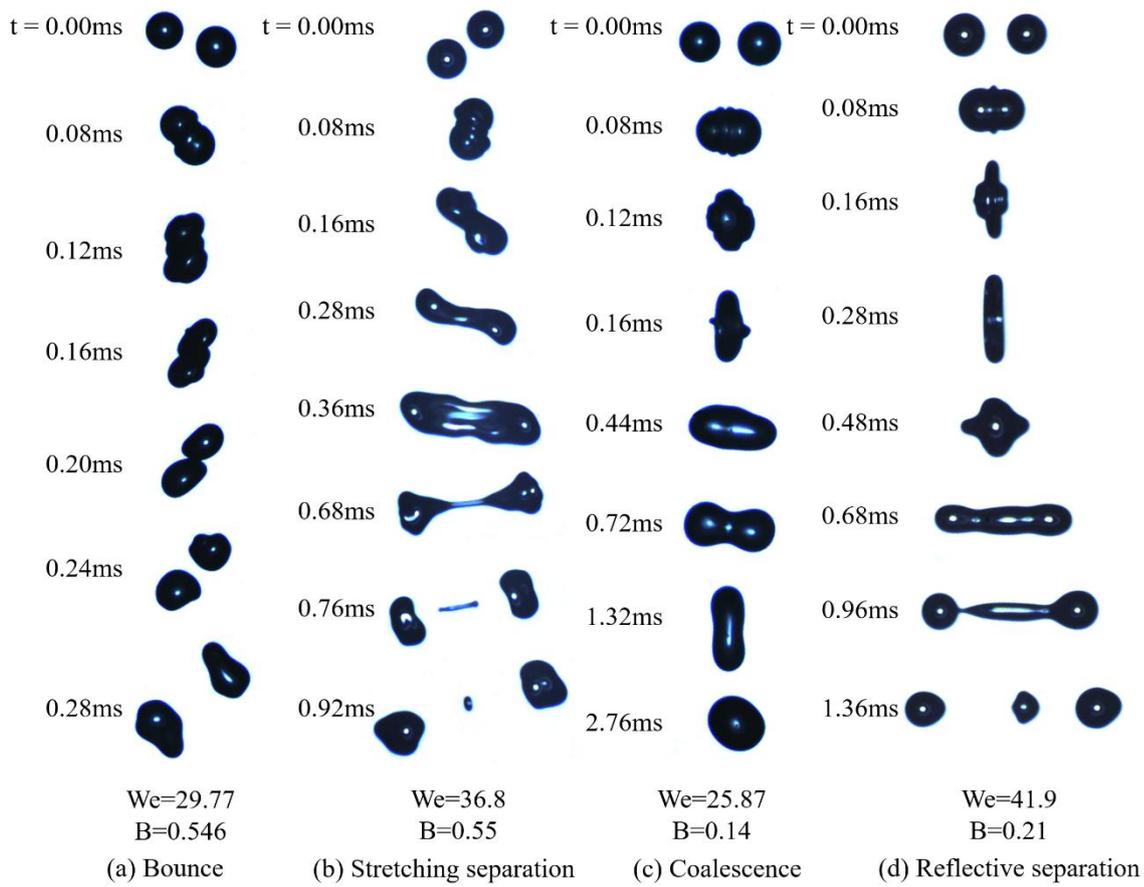

Figure S6. Different droplet collision outcomes under 21 atm, (a) Bouncing, (b) Stretching separation, (c) Coalescence, and (d) Reflective separation.

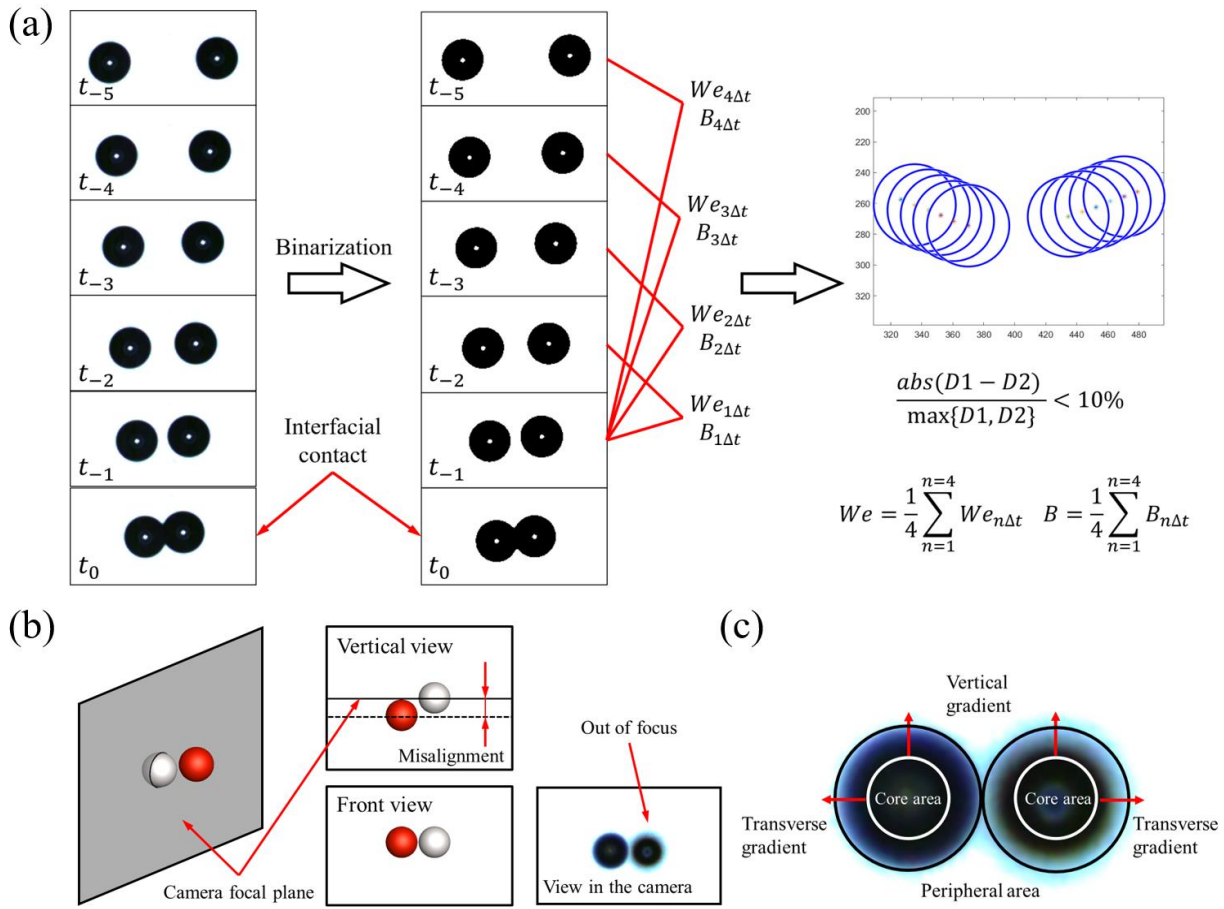

Figure S7. (a) Image processing process, (b) Possible misalignment during droplet collision in the high-pressure chamber, and (c) Definition of transverse and vertical gradients in judging droplet misalignment.

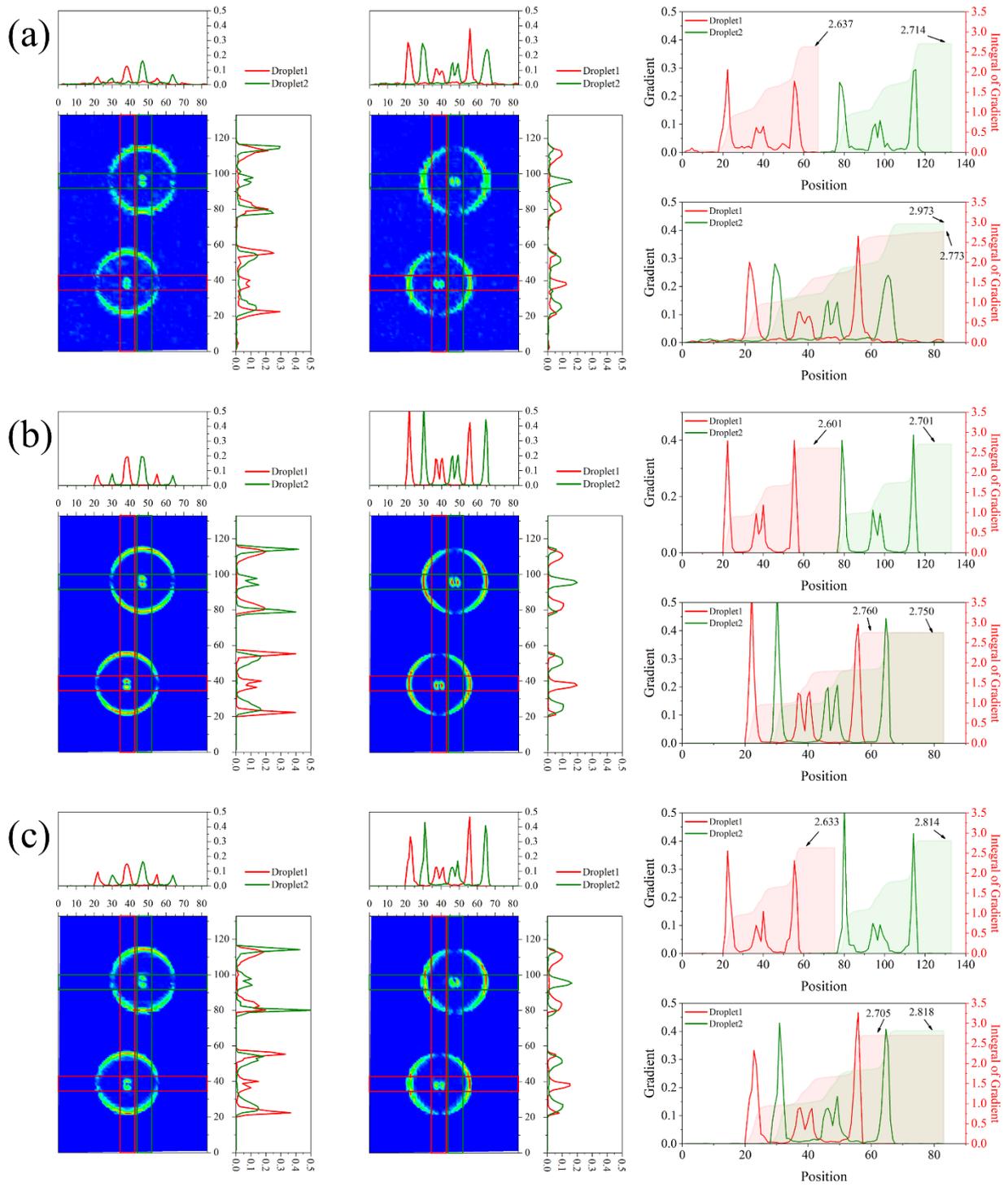

Figure S8. Droplet edges detection based on coalescence cases with different channels of (a) R, (b) G, and (c) B.

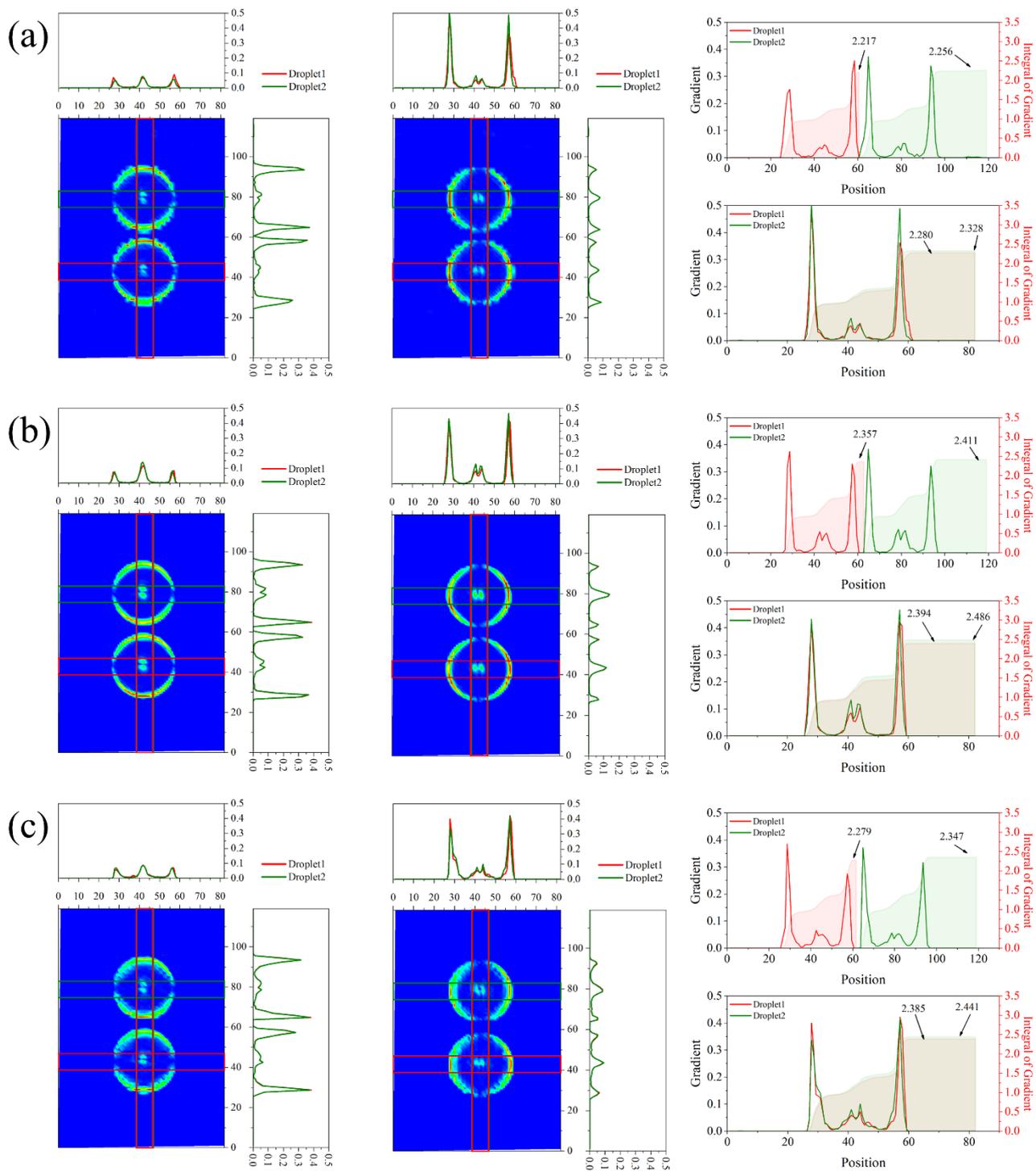

Figure S9. Droplet edges detection based on reflective separation cases with different channels of (a) R, (b) G, and (c) B.

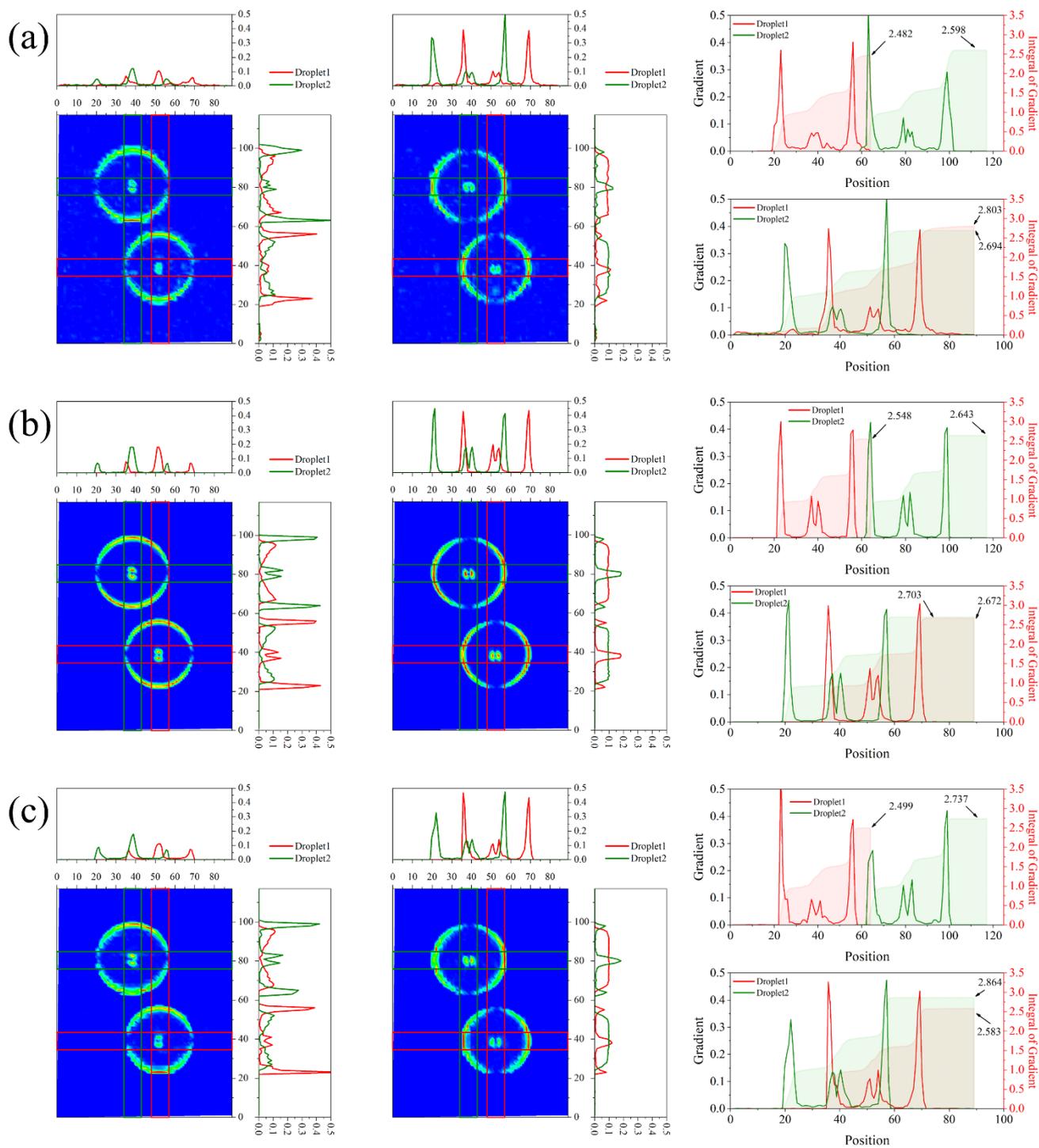

Figure S10. Droplet edges detection based on stretching separation cases with different channels of (a) R, (b) G, and (c) B.

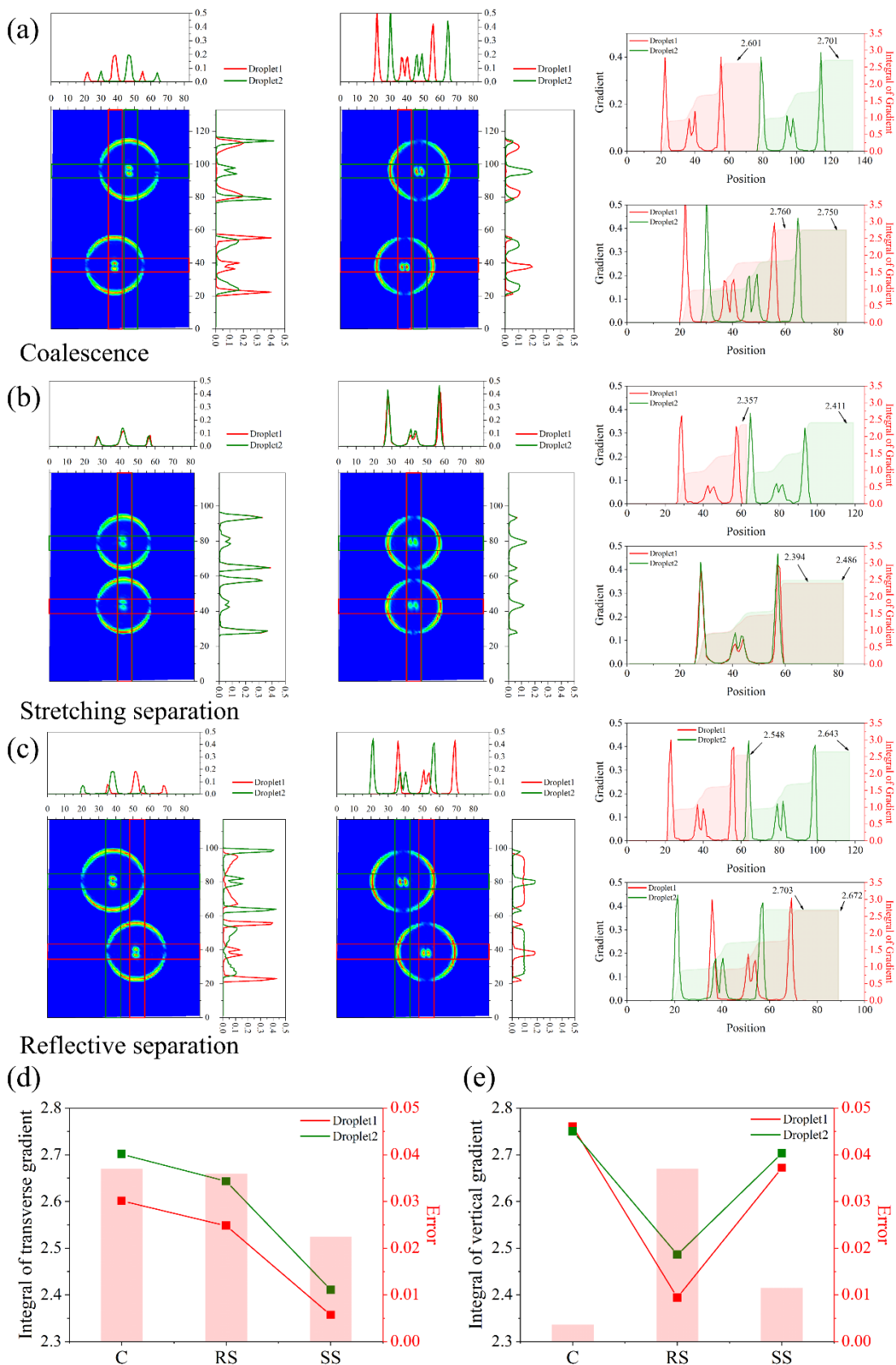

Figure S11. Method for identifying whether two droplets are misaligned in the direction of camera focal plane for (a) Coalescence, (b) Reflective separation, (c) Stretching separation, (d) Integral of transverse gradient, and (e) Integral of vertical gradient.

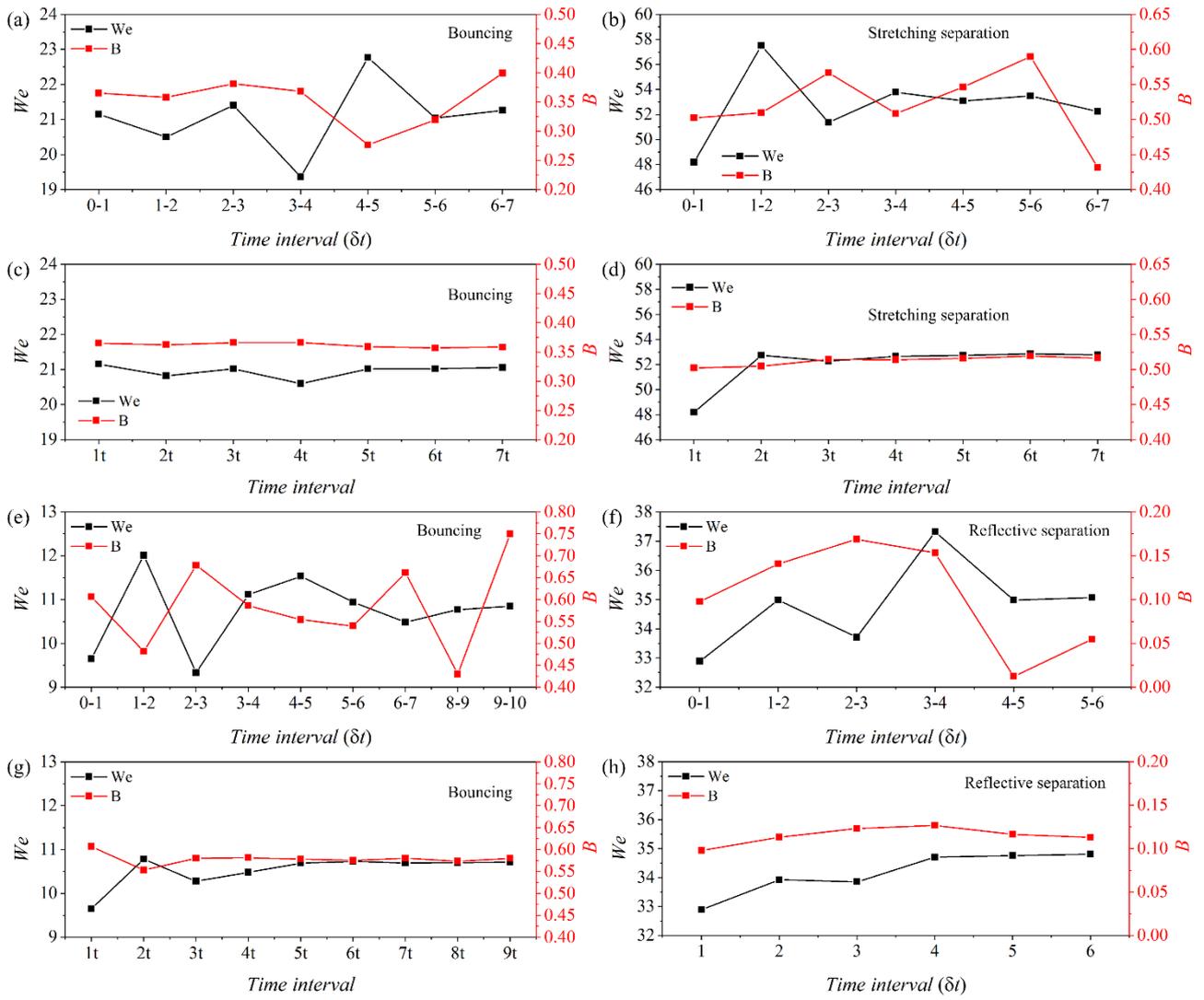

Figure S12. The determination of time intervals in data processing (a)-(d) 11 atm, (e)-(h) 41 atm.

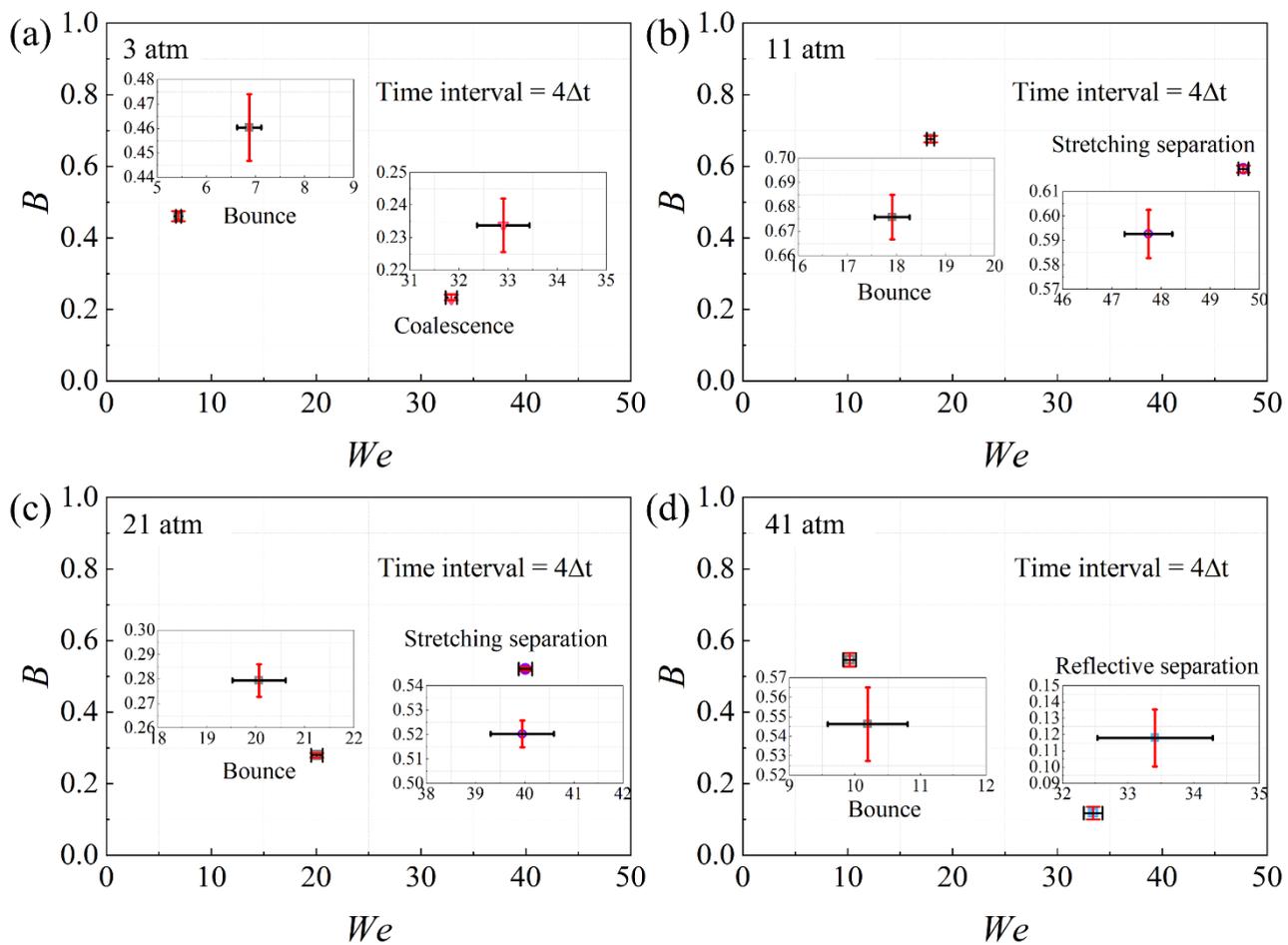

Figure S13. Error analysis of current data processing methods for (a) 3atm, (b) 11 atm, (c) 21 atm, and (d) 41 atm.

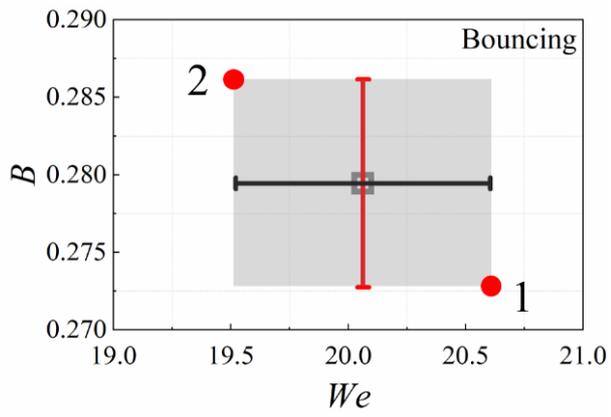

Figure S14. Different datasets of droplet collision utilized for the boundary determination taking droplet bouncing as an example.

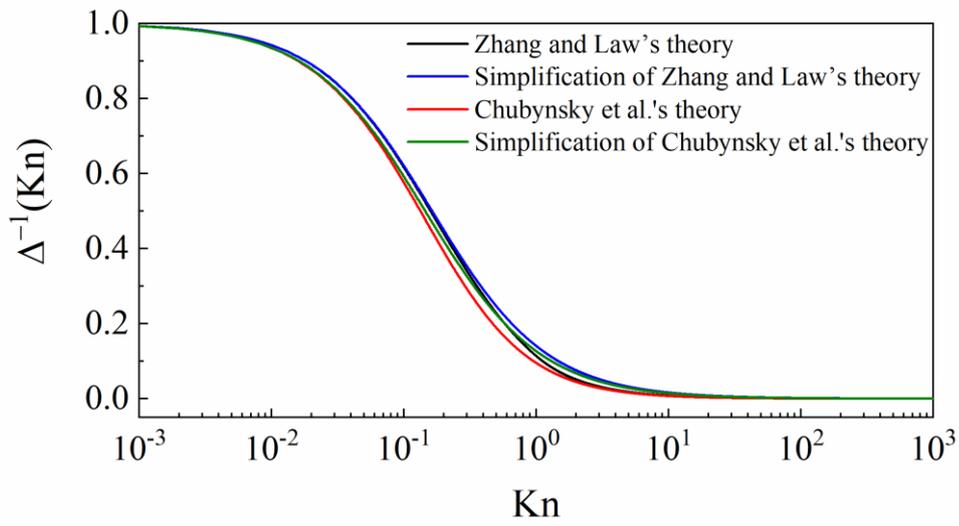

Figure S15. The comparison between Zhang and Law's theory and Chubynsky et al.'s theory and their simplified forms regarding the correction of the rarefied gas effect.